\documentclass[aps,prd,preprint,superscriptaddress]{revtex4-1}

\usepackage{latexsym}
\usepackage{amsmath,amssymb}
\usepackage{graphicx}
\usepackage{subfigure}
\usepackage{xcolor}
\usepackage{hyperref}     
\hypersetup{colorlinks,%
  citecolor=blue,%
  linkcolor=cyan,%
}

\begin{document}


\title{Thermodynamic phase transition based on the nonsingular temperature}

\author{Myungseok Eune}%
\email[]{eunems@smu.ac.kr}%
\affiliation{Department of Computer System Engineering, Sangmyung
  University, Cheonan, 330-720, Republic of Korea}%

\author{Yongwan Gim}%
\email[]{yongwan89@sogang.ac.kr}%
\affiliation{Department of Physics, Sogang University, Seoul 121-742,
  Republic of Korea}%

\author{Wontae Kim}%
\email[]{wtkim@sogang.ac.kr}%
\affiliation{Department of Physics, Sogang University, Seoul 121-742,
  Republic of Korea}%

\date{\today}

\begin{abstract}
The Hawking temperature for the Schwarzschild black hole
is divergent when the mass of the black hole vanishes; however,
the corresponding geometry becomes the Minkowski spacetime whose intrinsic
temperature is zero.
In connection with this issue,
we construct a nonsingular
temperature which follows the Hawking temperature for the
large black hole, while it vanishes when the black hole is completely
evaporated.
For the thermodynamic significances of this modified
temperature, we calculate thermodynamic quantities and
study phase transitions.
It turns out that even the small black hole can be stable below a certain temperature, and
the hot flat space is always metastable so that
it decays into the stable small black hole or the stable large black hole.
 \end{abstract}


\maketitle

\section{Introduction}
\label{sec:intro}

Bekenstein has suggested that a black hole should have an entropy
which is proportional to the area of the horizon~\cite{Bekenstein:1972tm,
  Bekenstein:1973ur, Bekenstein:1974ax}, and Hawking has shown that
there is radiation from the black hole through the analysis for
the origin of the entropy from the point of view of quantum field
theory~\cite{Hawking:1974sw}.  The Hawking temperature could
be defined generically as $T_{\rm H} =  \hbar \kappa_H /(2\pi)$, where
$\kappa_H$ is the surface gravity at the
horizon. Subsequently, the thermodynamics of the black hole based on this temperature has
been one of the most important issues in black hole physics \cite{Gross:1982cv, Hawking:1982dh, York:1986it},
so that there have been much intensive study of thermodynamics and phase transitions in various
black holes \cite{ Medved:1998ks, Cai:2001dz, Cai:2007vv, Cai:2007wz, Clement:2007tw, Banerjee:2008cf,  Kim:2008bf, Cai:2009ua, Myung:2010dv, Banerjee:2010da, Banerjee:2011au, Jardim:2012se, Rodrigues:2012dk, Eune:2013qs, Son:2012vj, Biro:2013cra, Gim:2014ira, Gim:2014nba}.
Such phase transitions could be easily read off from the behaviors of
the heat capacity and the
free energy.
In particular, considering the Schwarzschild black hole in a cavity
which defines the isothermal surface~\cite{York:1986it},
it was shown that the hot flat space is more probable
than the large black hole below a critical temperature,
while the large black hole is more probable than the hot flat space above
the critical temperature.
In fact, the most essential ingredient in the thermodynamics of black holes
is to define the black hole temperature
such as the Hawking
temperature for the Schwarzschild black hole given as $T_{\rm H}
=\hbar/ (8\pi G M)$ from the surface gravity,
where $G$ and $M$ are the gravitational constant
and the mass of the black hole, respectively.
However, it shows that the temperature is proportional to the inverse of the mass, and
it is divergent when the mass of the black hole vanishes, although the black hole
disappears and its metric becomes the Minkowski spacetime.

The conventional method to resolve
the above singular behavior of the Hawking temperature for the Schwarzschild black
hole is to introduce the Planck mass as a cutoff $M_{\rm P}$
in the regime of the generalized uncertainty principle (GUP)~\cite{Padmanabhan:1987au,
  Konishi:1989wk, Maggiore:1993kv, Maggiore:1993zu, Garay:1994en,
  Kempf:1994su, Chang:2001bm}, and then the GUP temperature is obtained as
$  T_{\rm GUP} = M/(4\pi) [1 \pm \sqrt{1 - M_{\rm P}^2/M^2 } ]$  \cite{Adler:2001vs}.
It produces the well-known Hawking temperature for the large black hole, and
it is finite at $M=M_{\rm  P}$.
Thanks to this cutoff,
the Hawking temperature can be regular as long as  $M \ge M_{\rm P}$; however, it requires
the remnant which has a nonvanishing temperature at the order of the Planck mass.
Moreover, the stability of the remnant is not warranted since the heat capacity of the remnant
approaches negative zero \cite{Kim:2007hf}.
There have been extensive studies of the thermodynamics of black holes
using the GUP temperature,
\cite{Custodio:2003jp, Medved:2004yu, Myung:2006qr,Kim:2007bx},
 and the discussion for the above issues of GUP by deforming the Einstein-Hilbert action which leads to the formation of a zero temperature stable remnant \cite{Isi:2013cxa}.

On the other hand, 
there has been another nontrivial attempt to derive the entropy and
temperature which give eventually
finite results when the mass of the black hole goes to zero
by defining a new temperature based on the argument of the high order of quantum corrections \cite{Singleton:2010gz, Singleton:2013ama}.
It turns out that the entropy goes to zero, while the
temperature is still divergent when the black hole evaporates completely.
Note that one parameter in the temperature is
replaced by a mass-dependent one, so that
it is possible to make the temperature finite as was shown in Ref. \cite{Singleton:2010gz};
however, it is still nonvanishing when the mass of the black hole
goes to zero. Now, one might wonder how to get the entropy and
temperature which follow the well-known ordinary Hawking temperature and entropy
and the vanishing temperature and entropy at the end state of
the black hole without any remnants.

In this work,
we would like to present a nonsingular temperature
without resort to the cutoff in the UV region.
Of course, the complete quantum gravity beyond the Planck scale is not yet known,
but for our purpose, we just demand two boundary
conditions - that the
temperature should follow the ordinary Hawking temperature for the large black hole
and vanish when the mass of the black hole becomes zero.
The latter condition is plausible if
the end state of evaporation of the black hole naturally becomes
the Minkowski spacetime without the remnant.
In Sec.~\ref{sec:tem},
we start with a somewhat general finite temperature, satisfying the above two
boundary conditions.
Apart from these conditions, we assume that
the entropy of the black hole is positive,
and it also vanishes when the mass of the black hole goes to zero.
At last, we determine
the most simple but regular temperature satisfying the
desired asymptotic behaviors among infinite number of ways to reach the Minkowskian spacetime.
In Sec.~\ref{sec:therm}, in order to exhibit the
thermodynamic behaviors of the temperature,
the above nonsingular temperature will be localized
in the cavity following the work in Ref. \cite{York:1986it}. Then,
we find that the localized temperature has two extrema, so that there
exist two critical masses  of $M_0$ and $M_1$ compared to the
single critical mass $M_1$ in the standard calculation in the cavity.
The conventional thermodynamic calculation showed
that the small black hole is always unstable; however,
in this case
it will be shown that
the small black hole can also be stable.
Next, in order for studying phase transition based on
the newly defined temperature, the free energies will be considered
for the hot flat space, the small black hole, and the large black hole, respectively.
Then, we find a Hawking-Page-type phase transition
between the small black hole and the large black hole.
Additionally, we find that the hot flat space is always metastable,
and it decays into the small black hole or the large black hole.
Finally, the conclusion and discussion are given in Sec.~\ref{Discus}.

\section{Nonsingular temperature of the black hole}
\label{sec:tem}

In the Schwarzschild black hole, the Hawking temperature
becomes singular at $M \to 0$, so that
it is nontrivial task to describe
this region properly by means of the ordinary Hawking
temperature.
So, let us assume that the Hawking temperature
can be modified in such a way that
the temperature of the black hole
vanishes when the mass of the black hole goes to zero
while it follows the behavior of
the well-known Hawking temperature for the large black hole.
For this purpose, let us write temperature as
\begin{equation}
  T = \frac{1}{8\pi G M}t(M), \label{T::t:M}
\end{equation}
where $t(M)$ should be chosen in order to satisfy the two boundary
conditions mentioned earlier and a more or less general expression
in terms of polynomial expansion of the mass can be written as
\begin{equation}
  t(M)=\frac{\sum_{i=0}^n a_i
    \left(\frac{M}{M_{\rm P}}\right)^{1+\alpha_i}}{\sum_{i=0}^n b_i
    \left(\frac{M}{M_{\rm P}}\right)^{1+\beta_i}+C}, \label{t:M}
\end{equation}
where $t \sim M^{1+ \alpha_0}$ for $M \to 0$
and $t \sim O(1) $ for $M \to \infty$, and
$\alpha_i$, $\beta_i$, and $C$ are positive constants
with $\alpha_n = \beta_n$.
Additionally, $\alpha_i < \alpha_j$, $\beta_i < \beta_j$ for $i<j$,
and $a_n = b_n$. Note that
the similar ansatz for the temperature appeared with the same spirit 
in terms of the expansion of integer powers in Ref. \cite{Singleton:2010gz}; however,
any fractional powers are allowed in Eq. \eqref{t:M} without loss of 
generality.   
Compactly, the modified temperature for the black hole
can be rewritten as
\begin{equation}\label{T}
  T=\frac{1}{8\pi G M}\frac{\sum_{i=0}^n a_i
    \left(\frac{M}{M_{\rm P}}\right)^{1+\alpha_i}}{\sum_{i=0}^n b_i
    \left(\frac{M}{M_{\rm P}}\right)^{1+\beta_i}+C}.
\end{equation}
Then, the entropy calculated from the first law
of thermodynamics,
\begin{equation}
  S=\int \frac{dM}{T}, \label{S:1st}
\end{equation}
still respects the
Bekenstein-Hawking entropy for the large black hole as $S \sim 4 \pi G M^2$.
On the other hand, the
entropy~\eqref{S:1st} for the small mass  can be calculated as
\begin{align}
\label{5}
  S & = \frac{8\pi G M_{\rm P} C}{a_0}\int \left( \frac{M_{\rm
        P}}{M} \right)^{\alpha_0} dM \sim \left\{\begin{array}{ll}
      -\frac{1}{M^{\alpha_0-1}} \qquad & \text{for}~ \alpha_0 > 1 , \\
      \ln M   \qquad & \text{for}~\alpha_0 = 1, \\
      M^{1-\alpha_0} \qquad & \text{for}~0<\alpha_0 < 1, \end{array}
  \right.
\end{align}
where we neglected the subleading terms.
Note that the entropy is negative divergent
for $\alpha_0 \ge 1$, and it vanishes for $0<\alpha_0<1$
when $M \to 0$.  As a result, we obtain the additional condition of
$0<\alpha_0<1$ in order for the positive entropy.

\begin{figure}[pt]
  \begin{center}
  \includegraphics[width=0.45\textwidth]{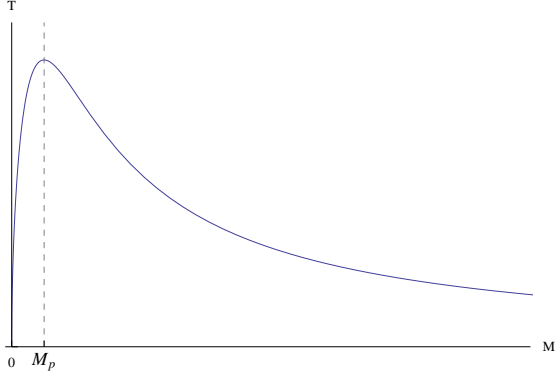}
  \end{center}
  \caption{The modified temperature is plotted by
    setting the constants as $M_{\rm P} = 1$ and $\alpha=1/2$ in Eq. \eqref{T:simple}.
    The temperature follows the Hawking temperature asymptotically and vanishes
    at $M \to 0$. The maximum temperature appears at the Planck mass. }
  \label{fig:T}
\end{figure}

The most simple form of the
modified temperature~\eqref{T} without loss of generality
corresponds to $n=0$, which is written as
\begin{equation}
  T=\frac{1}{8\pi G M}
  \left[1+\frac{1}{\alpha}\left(\frac{M_{\rm
          P}}{M}\right)^{1+\alpha}\right]^{-1},  \label{T:simple}
\end{equation}
where we used the relation of $M_{\rm P}^2 =
G^{-1}$, $\alpha_0 =\beta_0$, $a_0=b_0$, and in particular
chose $a_0 C^{-1} = \alpha_0$ in order to make the temperature
a maximum at $M = M_{\rm P}$.
After calculations, $\alpha_0$ was replaced by $\alpha$ for simplicity, so that
$0<\alpha<1$ in Eq. \eqref{T:simple}.
The behavior of the temperature~\eqref{T:simple} is illustrated in
Fig.~\ref{fig:T} which shows the modified temperature following
the Hawking temperature asymptotically and eventually vanishing when
the mass of the black hole goes to zero.
So, eventually, there is no remnant
after complete evaporation of the black hole, and the corresponding
geometry becomes the Minkowski spacetime.
It is compatible with the fact that the Minkowski spacetime
has no intrinsic temperature.

By using Eq.~\eqref{S:1st}, the entropy
corresponding to the modified temperature \eqref{T:simple} is
calculated as
\begin{equation}
  S=4\pi G  M^2 + \frac{8\pi}{\alpha(1-\alpha)} \left(\frac{M}{M_{\rm
        P}}\right)^{1-\alpha}, \label{S:simple}
\end{equation}
where the entropy vanishes for $M \to 0$. It goes to the Bekenstein-Hawking's area law of the entropy for the large black hole.
As a result, the temperature and entropy vanish at the end state of the
black hole. The difference from the previous works \cite{Singleton:2010gz, Singleton:2013ama} comes from the fact that even the temperature can be zero
when the mass of the black hole goes to zero if we use the fractional power of
expansion in the temperature expression \eqref{T:simple}.

\section{Thermodynamic quantities and phase transition}
\label{sec:therm}

\begin{figure}[pt]
  \begin{center}
    \subfigure[{local temperature}]{
      \includegraphics[width=0.45\textwidth]{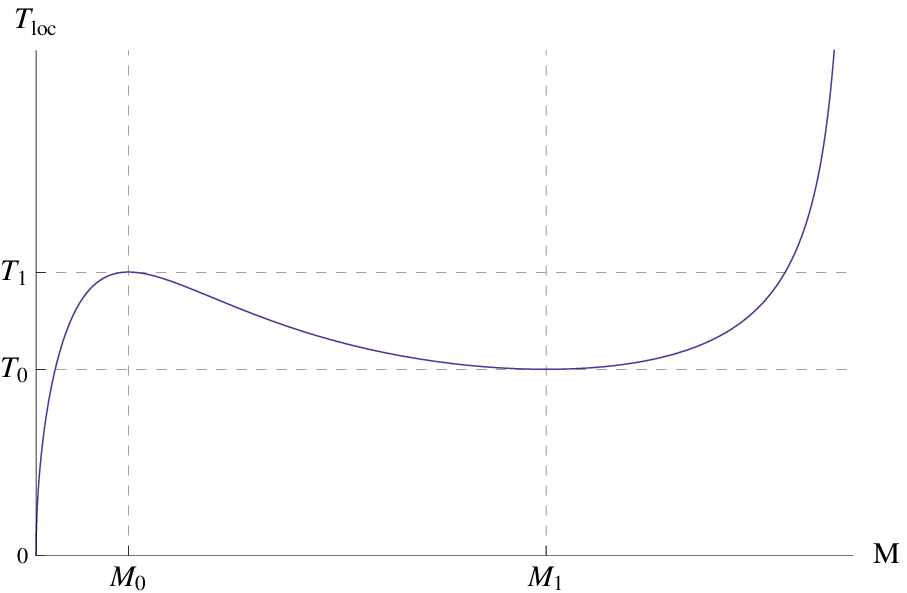}\label{fig:locT}}
    \subfigure[{heat capacity}]{
      \includegraphics[width=0.45\textwidth]{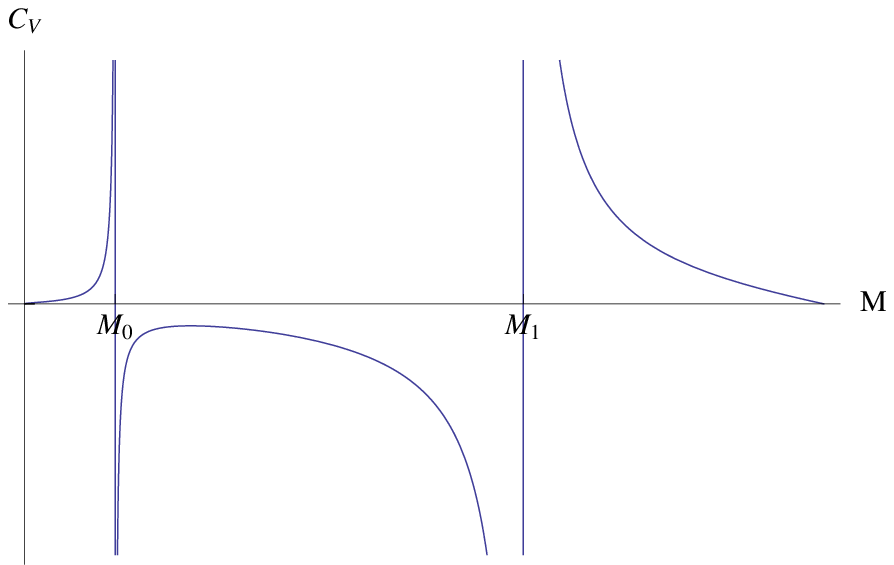}\label{fig:cap}}
  \end{center}
  \caption{The local temperature (a) and the heat capacity (b) are
    plotted by setting the constants as $G=1$,
    $\alpha=1/2$, and $r=10$, respectively.
   Figure (a) shows that the number of states of the black hole
    is subject to the temperature.
   In Fig. (b), the large
    black hole in $M>M_1$ is always stable, but the small black hole is stable for $M<M_0$
    whereas it is unstable for $M_0 < M<M_1$.}
  \label{fig:locTandC}
\end{figure}

In connection with the modified temperature \eqref{T:simple} ,
we are going to calculate the relevant thermodynamic quantities
which will be employed in order to investigate thermodynamic phase transitions.
Let us first consider the cavity as a boundary with a radius $r$ to
study quasilocal thermodynamics along the line of the procedure in Ref. \cite{York:1986it}.
Then, the local temperature measured
at the boundary is given as
\begin{align}
  T_{\rm loc} &= \frac{T}{\sqrt{1-2GM/r}} \notag \\
  &=\frac{1}{8\pi G M \sqrt{1-2GM/r}}
  \left[1+\frac{1}{\alpha}\left(\frac{M_{\rm P}}{M}\right)^{1+\alpha}\right]^{-1}, \label{T:local}
\end{align}
which is shown in Fig.~\ref{fig:locT}.
It shows that there are two extrema: one is
the local maximum at $M=M_0$ and the other is the local minimum at
$M=M_1$.
Let us now define the large black hole
where $M > M_1$ and the small black hole where $M<M_1$ for convenience.
From Fig.~\ref{fig:locT}, it can be shown that there is one small black hole for
$0<T < T_0$,  two small black holes  and one large black hole $T_0 < T < T_1$,
and one large black hole for $T>T_1$, respectively.
The similar black hole states can be found in the noncommutative black hole \cite{Nicolini:2005vd, Myung:2006mz, Nicolini:2008aj, Kim:2008vi}
and the Horava-Lifshitz black hole \cite{Cai:2009qs, Myung:2009dc, Eune:2012np};
however, the remnant associated with the parameter which characterizes the model
was required.

Note that the entropy \eqref{S:simple} is
independent of the size of the cavity in quasilocal thermodynamics \cite{Brown:1994gs}
since the degrees of freedom of the black hole should be independent
of the observer located at the radial distance $r$ \cite{Kim:2013zha}.
Applying the first law of thermodynamics, the total
thermodynamic internal energy within the boundary $r$ is obtained as
\begin{align}
E_{\rm loc} &= \int^M_0 T_{\rm loc} dS \notag \\
&= \frac{r}{G}-\frac{r}{G}\sqrt{1-\frac{2GM}{r}}, \label{E}
\end{align}
which is the same with the energy obtained from the conventional
Hawking temperature~\cite{York:1986it,Whiting:1988qr}.
Using Eq. \eqref{E}, the heat
capacity at fixed $r$  is calculated as
\begin{align}
  C_V &= \left(\frac{\partial E_{\rm loc}}{\partial T_{\rm loc}} \right)_r \notag\\
  &=\frac{8\pi G M^2\left(1-\frac{2G
        M}{r}\right)\left(1+\frac{1}{\alpha}\left(\frac{M_{\rm
            P}}{M}\right)^{1+\alpha}\right)}{\left(\frac{M_{\rm
          P}}{M}\right)^{1+\alpha}\left(1-\frac{2GM}{r}+\frac{1}{\alpha}\frac{G
        M}{r}\right)-1+\frac{3GM}{r}}.
\end{align}
As seen from Fig.~\ref{fig:cap}, the large black hole in $M> M_1$ is stable since the heat capacity is
positive, which is the same with the ordinary behavior of the heat capacity in
the Schwarzschild black hole. However,
the small black hole can be either stable or unstable
depending on the size of the black hole, so that
it is stable for $M < M_0$ and unstable for $M_0 < M < M_1$.
After all, it turns out that the small black hole for $M < M_0$
can be nucleated like the large black hole,
which makes the phase transition nontrivial.

\begin{figure}[pt]
  \begin{center}
 \subfigure[{\ Free energy based on the Hawking temperature}]{
   \includegraphics[width=0.45\textwidth]{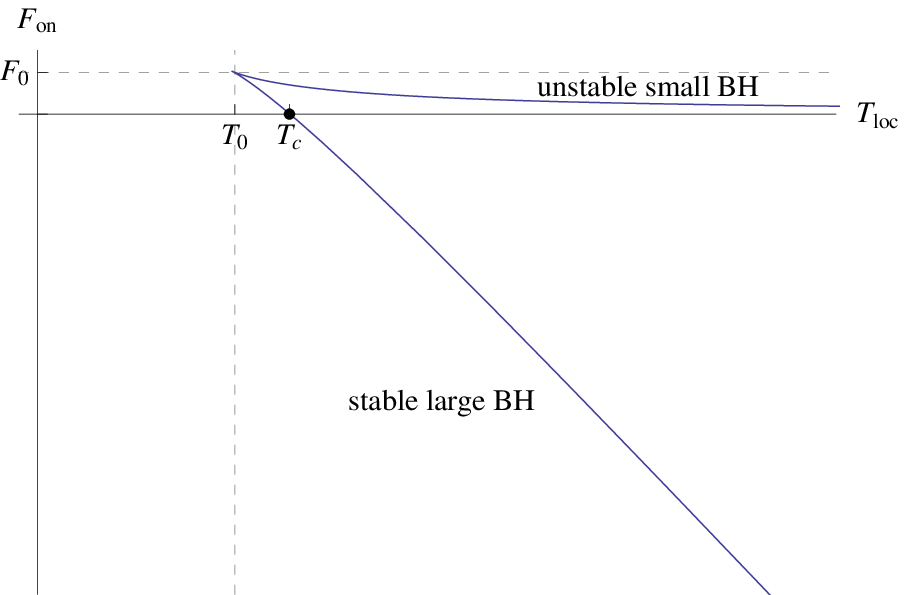} \label{fig:F:Sch}}
 \subfigure[{\ Free energy based on the modified temperature}]{
  \includegraphics[width=0.45\textwidth]{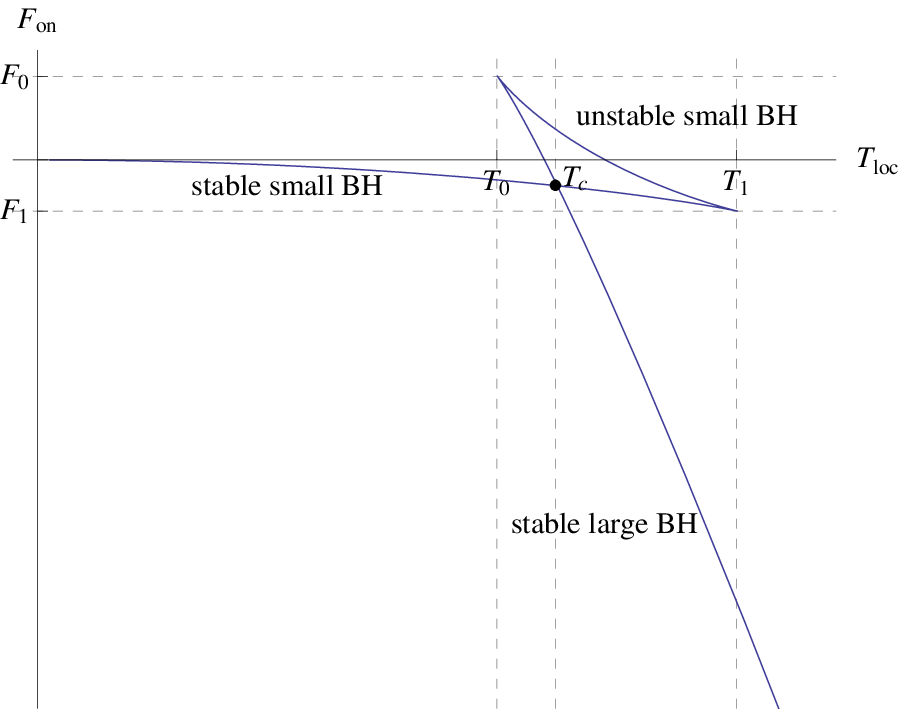} \label{fig:F:Sch:mod.T}}
  \end{center}
  \caption{The ordinary free energy and the modified free energy are plotted,
           respectively. The constants are chosen as $G=1$,  $\alpha=1/2$, and $r=10$.
   The phase transition
    appears between the hot flat space and the stable large black hole  in Fig. \ref{fig:F:Sch},
    while it happens between the stable small
    black hole and the stable large black hole in Fig. \ref{fig:F:Sch:mod.T}. $T_c$ is a
critical temperature and the other temperatures in Fig. \ref{fig:F:Sch:mod.T}
were defined in Fig.~\ref{fig:locT}.}
  \label{fig:F}
\end{figure}

Next, let us calculate the free energy of the black hole in order to
study phase transition between the black holes and the hot flat space.
Using Eqs. \eqref{S:simple},  \eqref{T:local}, and~\eqref{E},
the free energy of the black hole is easily obtained as
\begin{align}
  F_{\rm on}^{\rm bh}  &= E_{\rm loc} - T_{\rm loc} S \notag \\
  &= \frac{r}{G}-\frac{r}{G}\sqrt{1-\frac{2GM}{r}}-\frac{4\pi G M^2
    +\frac{8\pi G M_{\rm P}^2}{\alpha(1-\alpha)}\left(\frac{M}{M_{\rm
          P}}\right)^{1-\alpha}}{8\pi G M \sqrt{1-\frac{2 GM}{r}}
    \left(1+\frac{1}{\alpha}\left(\frac{M_{\rm
            P}}{M}\right)^{1+\alpha}\right)}. \label{F:on}
\end{align}
Note that for the limit of $M_{\rm P}/M \to 0$,
Eq. \eqref{F:on} is reduced to the well-known free energy of the black hole
in the conventional thermodynamics of the Schwarzschild black hole \cite{York:1986it},
and the free energy of the hot flat space also becomes zero, $F_{\rm on}^{\rm hfs}=0$.
In this case,  as seen from Fig. \ref{fig:F:Sch}, the large black hole
is more probable than the hot flat space above the critical temperature $T_c$,
and the hot flat space is more probable below the critical temperature $T_c$.

We are now in a position to discuss the free energy \eqref{F:on} explicitly.
The behavior of  the stable large black hole in Fig. \ref{fig:F:Sch:mod.T} is essentially the same as
that of the conventional one in Fig. \ref{fig:F:Sch}.
The difference comes from  the behavior of the small black hole since
the unstable small black hole in Fig. \ref{fig:F:Sch:mod.T}  starts with
the positive free energy at $T_0$ but
is terminated with the negative free energy at $T_1$.
Note that the free energy of the unstable small black hole
is always positive in any temperature  in Fig. \ref{fig:F:Sch}.
Moreover, the free energy of the newly defined stable small black hole in Fig. \ref{fig:F:Sch:mod.T}
is always negative in $T<T_1$,
and the free energy of the hot flat space is still  $F_{\rm on}^{\rm hfs}=0$ since
the free energy \eqref{F:on} vanishes for $M \to 0$, so that
the free energy of the stable small black hole is lower than
the free energy of the hot flat space.
So,
the hot flat space can collapse
not only to the stable large black hole for $T > T_1$ but also
to the stable small black hole for $T <T_0$.
In particular, as for $T_0 <T< T_1$,
the Hawking-Page-type critical
phenomenon can appear between the stable small black hole and the stable large black hole,
which is compared to the conventional phase transition
between the hot flat space and the stable large black hole as seen from Fig. \ref{fig:F:Sch}.
Thus, the stable small
black hole is more probable for $T_0 < T < T_c$ and the stable large black hole
is more probable for $T_c <T < T_1 $.
Additionally, the hot flat space in $T_0 <T< T_1$ decays into the
stable small black hole or the stable large black hole eventually.

\section{Conclusion and Discussion}
\label{Discus}
When the Schwarzschild black hole completely evaporates,
the metric describing the system becomes the Minkowski spacetime,
although the Hawking temperature is divergent; however,
the Minkowski spacetime does not have
its own intrinsic temperature since there is no event horizon.
In this work, we considered the modified temperature for the Schwarzschild black hole
by requiring two conditions: that
it should follow the Hawking temperature for the large black hole
and vanish when the mass of the black hole goes to zero.
The modified temperature was designed in order to show
the regular behavior of the temperature of the black hole without
encountering any singularity when $M \to 0$.
On the other hand, the several constants in the temperature \eqref{T:simple}
imply that there are an infinite number of ways for the modified temperature
to reach the origin; however, we fixed the constants
in such a way that the maximum value of the temperature
appears at the Planck scale and fortunately only a single constant
remains unfixed.
So the simplest form of the
temperature could be obtained, but it shares the above two conditions.
By the use of this temperature,
thermodynamic quantities were calculated in order
to study thermodynamic stability and phase transition of the black hole
along the well-established procedure.
It turns out that there is a single state of
the stable small black hole for $T < T_0$ or the stable
large black hole for $T > T_1$.
For $T_0 <T<T_1$, there are three black hole states which consist of
the two
small black holes and  one large black hole.
Apart from
the existence of the stable small black hole,
the most interesting thing to be distinguished from the standard
thermodynamics is that
the flat space is no longer a stable state thermodynamically
in any temperature since it
should always decay
into the stable small black hole or the stable large black hole,
so that the final state becomes a black hole state.

Finally, 
the method for the present regular temperature 
requires an explanation about the limitations of the present approach 
since we assumed a certain modification of the 
Hawking temperature as a function of the black hole mass but did not 
discuss the origin of such a modification. 
First of all, the temperature \eqref{T:simple} derived from 
the polynomial expansion of the black hole mass is, indeed, not unique even in spite
of the plausibility of reproducing the conventional Hawking temperature; for example,
another type of temperature such as $T=1 /(8\pi G M)(1-e^{-k (M/M_p)^{1+\alpha}})$, where $k$ is an
arbitrary positive constant also satisfies
the two boundary conditions mentioned in Sec. II.
To fix the physically meaningful temperature uniquely and figure out what happens
at the end state of evaporation of the black hole, the complete theory of quantum gravity 
covering the trans-Planckian regime
should be defined. The second limitation of our approach is that
we employed the classical metric for the local Tolman temperature 
as seen from Eq. \eqref{T:local}, which is a temporary expedient.
 In particular, one can expect that such a modified temperature 
 \eqref{T:simple} comes from a change in the spacetime geometry, so that the local temperature
\eqref{T:local} changes accordingly.  
Note that modifications in the geometry could have a nontrivial effect 
on our analysis and an impact on the physics of 
small black holes. This was noticed in the GUP regime, which
yields the corresponding GUP temperature, and 
the classical geometry should be changed according to 
the modification of the uncertainty relation \cite{Isi:2013cxa}.  
Based on this fact, using the one-to-one correspondence between the GUP and the GUP temperature, 
a corresponding modified uncertainty relation, which is written as
$\Delta x \Delta p+(2\ell_p/\alpha) (2\ell_p/\Delta x)^\alpha \Delta p \geq 1$,  can also be
derived straightforwardly from Eq. \eqref{T:simple}.
This modified uncertainty relation will modify the classical geometry similar to the result in Ref. \cite{Isi:2013cxa} so that
the local temperature will be changed somehow near the horizon and,
consequently, the thermodynamic behaviors of small black holes may be different
from the the present results. This deserves further study, which we hope will appear in the future.   

\acknowledgments
We would like to thank E. J. Son for exciting discussions.
W.K. was supported by the National Research Foundation of Korea(NRF) Grant  No. 2014R1A2A1A11049571, funded by the Korean government(MSIP).


\bibliographystyle{JHEP}       

\bibliography{references}

\providecommand{\href}[2]{#2}\begingroup\raggedright\begin{thebibliography}{10}

\bibitem{Bekenstein:1972tm}
J.~D. Bekenstein, {\it {Black holes and the second law}},  {\em Lett. Nuovo
  Cim.} {\bf 4} (1972) 737--740.

\bibitem{Bekenstein:1973ur}
J.~D. Bekenstein, {\it {Black holes and entropy}},  {\em Phys. Rev. D} {\bf 7}
  (1973) 2333--2346.

\bibitem{Bekenstein:1974ax}
J.~D. Bekenstein, {\it {Generalized second law of thermodynamics in black hole
  physics}},  {\em Phys. Rev. D} {\bf 9} (1974) 3292--3300.

\bibitem{Hawking:1974sw}
S.~Hawking, {\it {Particle Creation by Black Holes}},  {\em Commun. Math.
  Phys.} {\bf 43} (1975) 199--220.

\bibitem{Gross:1982cv}
D.~Gross, M.~Perry, and L.~Yaffe, {\it {Instability of Flat Space at Finite
  Temperature}},  {\em Phys. Rev. D} {\bf 25} (1982) 330--355.

\bibitem{Hawking:1982dh}
S.~Hawking and D.~N. Page, {\it {Thermodynamics of Black Holes in anti-De
  Sitter Space}},  {\em Commun. Math. Phys.} {\bf 87} (1983) 577.

\bibitem{York:1986it}
J.~York, James~W., {\it {Black hole thermodynamics and the Euclidean Einstein
  action}},  {\em Phys. Rev. D} {\bf 33} (1986) 2092--2099.

\bibitem{Medved:1998ks}
A.~Medved and G.~Kunstatter, {\it {Hamiltonian thermodynamics of charged black
  holes}},  {\em Phys.Rev.} {\bf D59} (1999) 104005,
  [\href{http://xxx.lanl.gov/abs/hep-th/9811052}{{\tt hep-th/9811052}}].

\bibitem{Cai:2001dz}
R.-G. Cai, {\it {Gauss-Bonnet black holes in AdS spaces}},  {\em Phys. Rev. D}
  {\bf 65} (2002) 084014, [\href{http://xxx.lanl.gov/abs/hep-th/0109133}{{\tt
  hep-th/0109133}}].

\bibitem{Cai:2007vv}
R.-G. Cai, L.-M. Cao, and Y.-W. Sun, {\it {Hawking-Page Phase Transition of
  black Dp-branes and R-charged black holes with an IR Cutoff}},  {\em JHEP}
  {\bf 0711} (2007) 039, [\href{http://xxx.lanl.gov/abs/0709.3568}{{\tt
  arXiv:0709.3568}}].

\bibitem{Cai:2007wz}
R.-G. Cai, S.~P. Kim, and B.~Wang, {\it {Ricci flat black holes and
  Hawking-Page phase transition in Gauss-Bonnet gravity and dilaton gravity}},
  {\em Phys. Rev. D} {\bf 76} (2007) 024011,
  [\href{http://xxx.lanl.gov/abs/0705.2469}{{\tt arXiv:0705.2469}}].

\bibitem{Clement:2007tw}
G.~Clement, J.~Fabris, and G.~Marques, {\it {Hawking radiation of linear
  dilaton black holes}},  {\em Phys.Lett.} {\bf B651} (2007) 54--57,
  [\href{http://xxx.lanl.gov/abs/0704.0399}{{\tt arXiv:0704.0399}}].

\bibitem{Banerjee:2008cf}
R.~Banerjee and B.~R. Majhi, {\it {Quantum Tunneling Beyond Semiclassical
  Approximation}},  {\em JHEP} {\bf 0806} (2008) 095,
  [\href{http://xxx.lanl.gov/abs/0805.2220}{{\tt arXiv:0805.2220}}].

\bibitem{Kim:2008bf}
W.~Kim and E.~J. Son, {\it {Thermodynamics of warped AdS(3) black hole in the
  brick wall method}},  {\em Phys.Lett.} {\bf B673} (2009) 90--94,
  [\href{http://xxx.lanl.gov/abs/0812.0876}{{\tt arXiv:0812.0876}}].

\bibitem{Cai:2009ua}
R.-G. Cai, L.-M. Cao, and N.~Ohta, {\it {Black Holes in Gravity with Conformal
  Anomaly and Logarithmic Term in Black Hole Entropy}},  {\em JHEP} {\bf 1004}
  (2010) 082, [\href{http://xxx.lanl.gov/abs/0911.4379}{{\tt
  arXiv:0911.4379}}].

\bibitem{Myung:2010dv}
Y.~S. Myung, {\it {Lifshitz black holes in the Ho\v{r}ava-Lifshitz gravity}},
  {\em Phys. Lett. B} {\bf 690} (2010) 534--540,
  [\href{http://xxx.lanl.gov/abs/1002.4448}{{\tt arXiv:1002.4448}}].

\bibitem{Banerjee:2010da}
R.~Banerjee, S.~Ghosh, and D.~Roychowdhury, {\it {New type of phase transition
  in Reissner Nordstrom - AdS black hole and its thermodynamic geometry}},
  {\em Phys.Lett.} {\bf B696} (2011) 156--162,
  [\href{http://xxx.lanl.gov/abs/1008.2644}{{\tt arXiv:1008.2644}}].

\bibitem{Banerjee:2011au}
R.~Banerjee and D.~Roychowdhury, {\it {Thermodynamics of phase transition in
  higher dimensional AdS black holes}},  {\em JHEP} {\bf 1111} (2011) 004,
  [\href{http://xxx.lanl.gov/abs/1109.2433}{{\tt arXiv:1109.2433}}].

\bibitem{Jardim:2012se}
D.~F. Jardim, M.~E. Rodrigues, and M.~Houndjo, {\it {Thermodynamics of phantom
  Reissner-Nordstrom-AdS black hole}},  {\em Eur.Phys.J.Plus} {\bf 127} (2012)
  123, [\href{http://xxx.lanl.gov/abs/1202.2830}{{\tt arXiv:1202.2830}}].

\bibitem{Rodrigues:2012dk}
M.~E. Rodrigues, D.~F. Jardim, and S.~J. Houndjo, {\it {Thermodynamics of black
  plane solution}},  {\em Gen.Rel.Grav.} {\bf 45} (2013) 2309--2323,
  [\href{http://xxx.lanl.gov/abs/1205.3481}{{\tt arXiv:1205.3481}}].

\bibitem{Eune:2013qs}
M.~Eune, W.~Kim, and S.-H. Yi, {\it {Hawking-Page phase transition in BTZ black
  hole revisited}},  {\em JHEP} {\bf 1303} (2013) 020,
  [\href{http://xxx.lanl.gov/abs/1301.0395}{{\tt arXiv:1301.0395}}].

\bibitem{Son:2012vj}
E.~J. Son and W.~Kim, {\it {Two critical phenomena in the exactly soluble
  quantized Schwarzschild black hole}},  {\em JHEP} {\bf 1303} (2013) 060,
  [\href{http://xxx.lanl.gov/abs/1212.2307}{{\tt arXiv:1212.2307}}].

\bibitem{Biro:2013cra}
T.~S. Biró and V.~G. Czinner, {\it {A $q$-parameter bound for particle spectra
  based on black hole thermodynamics with Rényi entropy}},  {\em Phys.Lett.}
  {\bf B726} (2013) 861--865, [\href{http://xxx.lanl.gov/abs/1309.4261}{{\tt
  arXiv:1309.4261}}].

\bibitem{Gim:2014ira}
Y.~Gim and W.~Kim, {\it {Thermodynamic phase transition in the rainbow
  Schwarzschild black hole}},  {\em JCAP} {\bf 1410} (2014) 003,
  [\href{http://xxx.lanl.gov/abs/1406.6475}{{\tt arXiv:1406.6475}}].

\bibitem{Gim:2014nba}
Y.~Gim, W.~Kim, and S.-H. Yi, {\it {The first law of thermodynamics in Lifshitz
  black holes revisited}},  {\em JHEP} {\bf 1407} (2014) 002,
  [\href{http://xxx.lanl.gov/abs/1403.4704}{{\tt arXiv:1403.4704}}].

\bibitem{Padmanabhan:1987au}
T.~Padmanabhan, {\it {Limitations on the Operational Definition of Space-time
  Events and Quantum Gravity}},  {\em Class.Quant.Grav.} {\bf 4} (1987)
  L107--L113.

\bibitem{Konishi:1989wk}
K.~Konishi, G.~Paffuti, and P.~Provero, {\it {Minimum Physical Length and the
  Generalized Uncertainty Principle in String Theory}},  {\em Phys.Lett.} {\bf
  B234} (1990) 276.

\bibitem{Maggiore:1993kv}
M.~Maggiore, {\it {The Algebraic structure of the generalized uncertainty
  principle}},  {\em Phys.Lett.} {\bf B319} (1993) 83--86,
  [\href{http://xxx.lanl.gov/abs/hep-th/9309034}{{\tt hep-th/9309034}}].

\bibitem{Maggiore:1993zu}
M.~Maggiore, {\it {Quantum groups, gravity and the generalized uncertainty
  principle}},  {\em Phys.Rev.} {\bf D49} (1994) 5182--5187,
  [\href{http://xxx.lanl.gov/abs/hep-th/9305163}{{\tt hep-th/9305163}}].

\bibitem{Garay:1994en}
L.~J. Garay, {\it {Quantum gravity and minimum length}},  {\em Int.J.Mod.Phys.}
  {\bf A10} (1995) 145--166, [\href{http://xxx.lanl.gov/abs/gr-qc/9403008}{{\tt
  gr-qc/9403008}}].

\bibitem{Kempf:1994su}
A.~Kempf, G.~Mangano, and R.~B. Mann, {\it {Hilbert space representation of the
  minimal length uncertainty relation}},  {\em Phys.Rev.} {\bf D52} (1995)
  1108--1118, [\href{http://xxx.lanl.gov/abs/hep-th/9412167}{{\tt
  hep-th/9412167}}].

\bibitem{Chang:2001bm}
L.~N. Chang, D.~Minic, N.~Okamura, and T.~Takeuchi, {\it {The Effect of the
  minimal length uncertainty relation on the density of states and the
  cosmological constant problem}},  {\em Phys.Rev.} {\bf D65} (2002) 125028,
  [\href{http://xxx.lanl.gov/abs/hep-th/0201017}{{\tt hep-th/0201017}}].

\bibitem{Adler:2001vs}
R.~J. Adler, P.~Chen, and D.~I. Santiago, {\it {The Generalized uncertainty
  principle and black hole remnants}},  {\em Gen.Rel.Grav.} {\bf 33} (2001)
  2101--2108, [\href{http://xxx.lanl.gov/abs/gr-qc/0106080}{{\tt
  gr-qc/0106080}}].

\bibitem{Kim:2007hf}
W.~Kim, E.~J. Son, and M.~Yoon, {\it {Thermodynamics of a black hole based on a
  generalized uncertainty principle}},  {\em JHEP} {\bf 0801} (2008) 035,
  [\href{http://xxx.lanl.gov/abs/0711.0786}{{\tt arXiv:0711.0786}}].

\bibitem{Custodio:2003jp}
P.~S. Custodio and J.~Horvath, {\it {The Generalized uncertainty principle,
  entropy bounds and black hole (non)evaporation in a thermal bath}},  {\em
  Class.Quant.Grav.} {\bf 20} (2003) L197--L203,
  [\href{http://xxx.lanl.gov/abs/gr-qc/0305022}{{\tt gr-qc/0305022}}].

\bibitem{Medved:2004yu}
A.~Medved and E.~C. Vagenas, {\it {When conceptual worlds collide: The GUP and
  the BH entropy}},  {\em Phys. Rev. D} {\bf 70} (2004) 124021,
  [\href{http://xxx.lanl.gov/abs/hep-th/0411022}{{\tt hep-th/0411022}}].

\bibitem{Myung:2006qr}
Y.~S. Myung, Y.-W. Kim, and Y.-J. Park, {\it {Black hole thermodynamics with
  generalized uncertainty principle}},  {\em Phys.Lett.} {\bf B645} (2007)
  393--397, [\href{http://xxx.lanl.gov/abs/gr-qc/0609031}{{\tt
  gr-qc/0609031}}].

\bibitem{Kim:2007bx}
W.~Kim and J.~J. Oh, {\it {Determining the minimal length scale of the
  generalized uncertainty principle from the entropy-area relationship}},  {\em
  JHEP} {\bf 0801} (2008) 034, [\href{http://xxx.lanl.gov/abs/0709.0581}{{\tt
  arXiv:0709.0581}}].

\bibitem{Isi:2013cxa}
M.~Isi, J.~Mureika, and P.~Nicolini, {\it {Self-Completeness and the
  Generalized Uncertainty Principle}},  {\em JHEP} {\bf 1311} (2013) 139,
  [\href{http://xxx.lanl.gov/abs/1310.8153}{{\tt arXiv:1310.8153}}].

\bibitem{Singleton:2010gz}
D.~Singleton, E.~C. Vagenas, T.~Zhu, and J.-R. Ren, {\it {Insights and possible
  resolution to the information loss paradox via the tunneling picture}},  {\em
  JHEP} {\bf 1008} (2010) 089, [\href{http://xxx.lanl.gov/abs/1005.3778}{{\tt
  arXiv:1005.3778}}].

\bibitem{Singleton:2013ama}
D.~Singleton, E.~C. Vagenas, and T.~Zhu, {\it {Self-similarity, conservation of
  entropy/bits and the black hole information puzzle}},  {\em JHEP} {\bf 1405}
  (2014) 074, [\href{http://xxx.lanl.gov/abs/1311.2015}{{\tt
  arXiv:1311.2015}}].

\bibitem{Nicolini:2005vd}
P.~Nicolini, A.~Smailagic, and E.~Spallucci, {\it {Noncommutative geometry
  inspired Schwarzschild black hole}},  {\em Phys.Lett.} {\bf B632} (2006)
  547--551, [\href{http://xxx.lanl.gov/abs/gr-qc/0510112}{{\tt
  gr-qc/0510112}}].

\bibitem{Myung:2006mz}
Y.~S. Myung, Y.-W. Kim, and Y.-J. Park, {\it {Thermodynamics and evaporation of
  the noncommutative black hole}},  {\em JHEP} {\bf 0702} (2007) 012,
  [\href{http://xxx.lanl.gov/abs/gr-qc/0611130}{{\tt gr-qc/0611130}}].

\bibitem{Nicolini:2008aj}
P.~Nicolini, {\it {Noncommutative Black Holes, The Final Appeal To Quantum
  Gravity: A Review}},  {\em Int.J.Mod.Phys.} {\bf A24} (2009) 1229--1308,
  [\href{http://xxx.lanl.gov/abs/0807.1939}{{\tt arXiv:0807.1939}}].

\bibitem{Kim:2008vi}
W.~Kim, E.~J. Son, and M.~Yoon, {\it {Thermodynamic similarity between the
  noncommutative Schwarzschild black hole and the Reissner-Nordstrom black
  hole}},  {\em JHEP} {\bf 0804} (2008) 042,
  [\href{http://xxx.lanl.gov/abs/0802.1757}{{\tt arXiv:0802.1757}}].

\bibitem{Cai:2009qs}
R.-G. Cai, L.-M. Cao, and N.~Ohta, {\it {Thermodynamics of Black Holes in
  Horava-Lifshitz Gravity}},  {\em Phys.Lett.} {\bf B679} (2009) 504--509,
  [\href{http://xxx.lanl.gov/abs/0905.0751}{{\tt arXiv:0905.0751}}].

\bibitem{Myung:2009dc}
Y.~S. Myung and Y.-W. Kim, {\it {Thermodynamics of Horava-Lifshitz black
  holes}},  {\em Eur.Phys.J.} {\bf C68} (2010) 265--270,
  [\href{http://xxx.lanl.gov/abs/0905.0179}{{\tt arXiv:0905.0179}}].

\bibitem{Eune:2012np}
M.~Eune, B.~Gwak, and W.~Kim, {\it {Local thermodynamics of KS black hole}},
  {\em Phys.Lett.} {\bf B718} (2013) 1505--1509,
  [\href{http://xxx.lanl.gov/abs/1209.4698}{{\tt arXiv:1209.4698}}].

\bibitem{Brown:1994gs}
J.~D. Brown, J.~Creighton, and R.~B. Mann, {\it {Temperature, energy and heat
  capacity of asymptotically anti-de Sitter black holes}},  {\em Phys.Rev.}
  {\bf D50} (1994) 6394--6403,
  [\href{http://xxx.lanl.gov/abs/gr-qc/9405007}{{\tt gr-qc/9405007}}].

\bibitem{Kim:2013zha}
W.~Kim, S.~Kulkarni, and S.-H. Yi, {\it {Quasilocal Conserved Charges in a
  Covariant Theory of Gravity}},  {\em Phys.Rev.Lett.} {\bf 111} (2013), no.~8
  081101, [\href{http://xxx.lanl.gov/abs/1306.2138}{{\tt arXiv:1306.2138}}].

\bibitem{Whiting:1988qr}
B.~F. Whiting and J.~York, James~W., {\it {Action Principle and Partition
  Function for the Gravitational Field in Black Hole Topologies}},  {\em
  Phys.Rev.Lett.} {\bf 61} (1988) 1336.

\end{thebibliography}\endgroup


%
%

\end{document}